\documentclass{appolb}%
\usepackage{epsfig}
\usepackage{rotating}
\usepackage{hyperref}
\usepackage{comment,color}
\usepackage{braket}
\usepackage{amssymb,amsmath,bm}
\usepackage{setspace,lipsum}
\usepackage{amsmath}
\usepackage{amsfonts}
\usepackage{amssymb}
\usepackage{graphicx}%
\begin{document}

\title{Time evolution of an unstable quantum system}
\author{F.~Giacosa$^{a,b}$
\address{$^a$Institute of Physics, Jan Kochanowski University, 25-406 Kielce, Poland}
\address{$^b$Institut f\"ur Theoretische Physik, Johann Wolfgang Goethe-
Universit\"at, 60438 Frankfurt am Main, Germany} }
\maketitle

\begin{abstract}
After reviewing the description of an unstable state in the framework of Lee
Hamiltonians (valid both for Quantum Mechanics (QM) and Quantum Field Theory
(QFT)), we consider some theoretical aspects of non-exponential decays: the
case of two decay channels, the broadening of the energy spectrum at short
times, the effect of an imperfect measurement, the link to QFT, and the decay
of an unstable moving particle with definite momentum. All the presented
effects were not confirmed in experiments, hence are at the present stage predictions.

\end{abstract}


\section{Introduction}

Decays of unstable states take place in quite different areas of physics,
which range from atomic and molecular phenomena (such as spontaneous emission)
up to elementary particles (such as the Higgs bosons). The survival
probability $p(t)$ for an unstable state is typically very well described by
an exponential function, $e^{-\Gamma t}$. Yet, it is nowadays well understood
both in Quantum Mechanics (QM) \cite{khalfin,ghirardi} and (at least partly)
in Quantum Field Theory (QFT) \cite{zenoqft,duecan} that $p(t)$ is not exactly
exponential: deviations at short as well as at long times appear. These
deviations were verified experimentally in Ref. \cite{raizen1} for short times
and in Ref. \cite{rothe} for long times (for an indirect evidence, see also
Ref. \cite{kelkar}). As a consequence of non-exponential decays, the famous
Quantum Zeno Effect (QZE) is realized when repeated ideal measurements are
performed at $\tau,$ $2\tau,$ etc. (usually called bang-bang measurements)
\cite{dega,misra,facchiprl,shimizu}: the survival probability approaches one
for $\tau\rightarrow0$. Experimentally, the QZE was measured by reducing the
probability of Rabi oscillations between atomic energy levels in Ref.
\cite{itano,balzer} and for a genuine unstable tunneling process in Ref.
\cite{raizen2}. Quite remarkably, as presented theoretically in Ref.
\cite{schulman} and verified experimentally in Ref. \cite{ketterle}, the QZE
takes place also for continuous measurements.

In this work, we first briefly review the mathematical treatment of an
unstable state through Lee Hamiltonians \cite{lee}. For definiteness, we
concentrate on a simple cutoff model in which two physical aspects, the
left-hand threshold and a cutoff at high energies, are simultaneously present.
This model nicely reproduces the purely exponential decay when the cutoff is
sent to infinity. Then, we discuss some interesting modern developments: (i)
The case of two (or more) decay channels \cite{duecan}. (ii)The final state
spectrum of a decay process such as spontaneous emission \cite{giacosapra}.
(iii) The QZE induced by an imperfect detector \cite{gplast}. (iv) The link to
QFT \cite{zenoqft}. (v) The decay of moving particle with a definite momentum
\cite{time}.

\section{Aspects of non-exponential decay}

\textbf{Lee Hamiltonian(s): }The Lee Hamiltonian $H=H_{0}+H_{1}$ couples an
unstable quantum state $\left\vert S\right\rangle $ to final states
$\left\vert k\right\rangle $ \cite{duecan,lee,giacosapra}:%
\begin{equation}
H_{0}=M_{0}\left\vert S\right\rangle \left\langle S\right\vert +\int_{-\infty
}^{+\infty}dk\omega(k)\left\vert k\right\rangle \left\langle k\right\vert
\text{ , }H_{1}=\int_{-\infty}^{+\infty}dk\frac{gf(k)}{\sqrt{2\pi}}\left(
\left\vert S\right\rangle \left\langle k\right\vert +\text{h.c.}\right)
\text{ .}%
\end{equation}
Usually, $\left\vert k\right\rangle $ represents a two-particle state emitted
back-to-back. The survival probability amplitude of the state $\left\vert
S\right\rangle $ reads%
\begin{equation}
a(t)=\left\langle S\right\vert e^{-iHt}\left\vert S\right\rangle \text{ }%
=\int_{-\infty}^{\infty}dmd_{S}(m)e^{-imt} \label{aoft}%
\end{equation}
where $d_{S}(m)$ is the energy distribution of the unstable state. The
survival probability is $p(t)=\left\vert a(t)\right\vert ^{2}.$ Here, we work
with a simplified model in which $\omega(k)=k$ and $f(k)=\theta(\Lambda
-k)\theta(k-E_{th})$ \cite{gplast} ($E_{th}<\Lambda$). In this way, the
unstable state $\left\vert S\right\rangle $ couples in a limited energy range
to the final states. In general, the outcome of the time evolution is
$e^{-iHt}\left\vert S\right\rangle =a(t)\left\vert S\right\rangle
+\int_{-\infty}^{+\infty}dkb(k,t)\left\vert k\right\rangle $.%
\begin{figure}
[ptb]
\begin{center}
\includegraphics[
height=1.8524in,
width=2.9507in
]%
{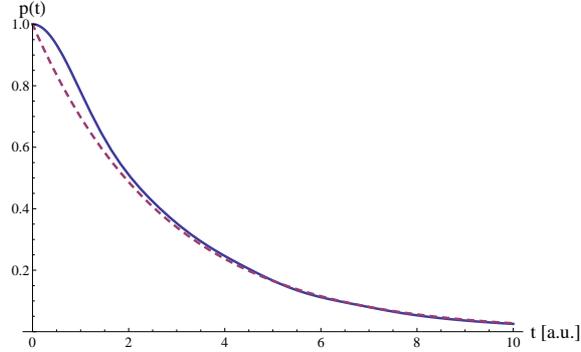}%
\caption{Survival probability $p(t)$ (solid line) in the cutoff model upon
using $E_{th}=0,$ $\Lambda=5,$ $M_{0}=3,$ $g^{2}=0.6^{2}$ in a.u. of the
energy. The dashed line refers to the corresponding exponential case,
$e^{-\Gamma t}$ with $\Gamma=g^{2}.$}%
\label{fig1}%
\end{center}
\end{figure}
When $E_{th}$ and $\Lambda$ are finite, deviations both at short and long
times occur, see the explicit numerical results in \cite{giacosapra} and, for
a particular illustrative numerical choice, Fig. 1. In the limit
$E_{th}\rightarrow-\infty,$ $\Lambda\rightarrow\infty$ the model reduces
exactly to the exponential decay \cite{giacosapra}: $a(t)=e^{-i(M_{0}%
-i\Gamma/2)t}$ (with $\Gamma=g^{2}$).

\textbf{Two decay channels \cite{duecan}:} The Lee Hamiltonian is easily
generalized to the case of two decay channels. In particular, we shall
consider $h_{1}(t)dt$ as the probability that $\left\vert S\right\rangle $
decays in the first channel between $t$ and $t+dt$ ($h_{2}(t)$ is the same
object in the second channel.) Then, it is useful to study the ratio
$R(t)=h_{1}(t)/h_{2}(t),$ which reduces to a constant $R(t)=\Gamma_{1}%
/\Gamma_{2}$ (ratio of decay widths) in the exponential limit. As shown in
Fig. 2, this ratio shows interesting fluctuations. Moreover, it deviates from
the constant limit for a quite long time (it does not flatten on it), thus it
is potentially interesting to be measured in future experiments.%

\begin{figure}
[ptb]
\begin{center}
\includegraphics[
height=2.0531in,
width=3.3771in
]%
{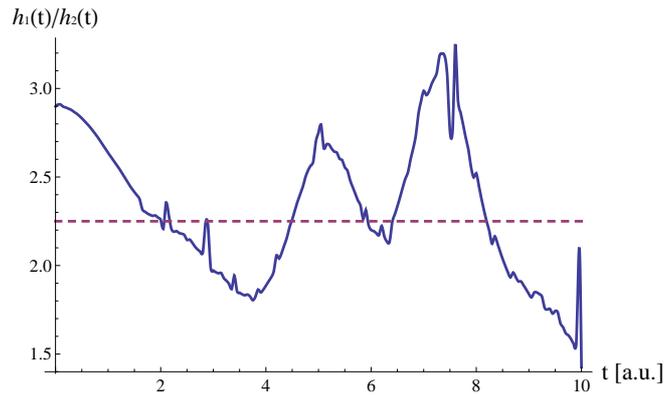}%
\caption{Ratio $R(t)=h_{1}(t)/h_{2}(t)$ (solid line) upon using $M_{0}%
=3,E_{th,1}=0,$ $\Lambda_{1}=5,$ $g_{1}^{2}=0.6^{2},$ $E_{th,2}=0.5,$
$\Lambda_{2}=4,$ $g_{2}^{2}=0.4^{2}$ in a.u. of the energy. The dashed line
refers to the exponential case, $\Gamma_{1}/\Gamma_{2}$ with $\Gamma_{k}%
=g_{k}^{2}.$}%
\end{center}
\end{figure}

\textbf{Energy spreading of the final state \cite{giacosapra}}: One studies
the function $\eta(t,\omega)$ defined as the probability that, by measuring
the final state at the time $t$, it has an energy between $\omega$ and
$\omega+d\omega.$ In the case of spontaneous photon emission, $\eta
(t,\omega)d\omega$ is the energy distribution of the photon at the time $t.$
In Fig. 3 we show this function for various values of $t.$ When $t$ is small,
this spectrum is large. Hence, if it could be possible to measure emitted
photons soon enough, they should show a larger spectrum than the simple decay
width $\Gamma.$%

\begin{figure}
[ptb]
\begin{center}
\includegraphics[
height=1.7945in,
width=3.3702in
]%
{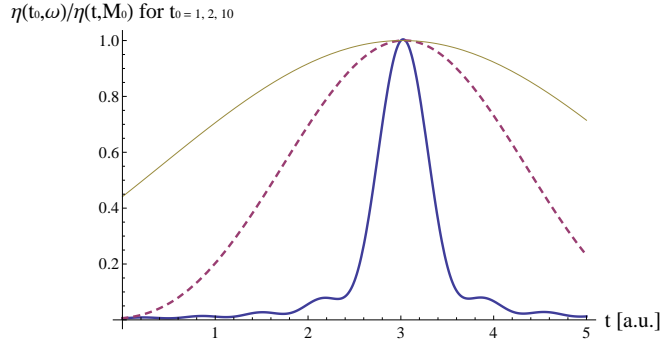}%
\caption{ $\eta(t_{0},\omega)/\eta(t_{0},M_{0})$ for the parameters of Fig. 1
and for $t_{0}=1$ (upper), $t_{0}=2$ (dashed), $t_{0}=10$ (lower curve). Note
the broadening for small $t_{0}$.}%
\end{center}
\end{figure}

\textbf{QZE induced by an imperfect measurements \cite{gplast}: }Next, we
assume that the detector can only detect the final state $\left\vert
k\right\rangle $ if $M_{0}-\lambda\leq k\leq M_{0}+\lambda.$ This means that
the probability to \textquotedblleft hear\textquotedblright\ the click of the
detector for a measurement at $\tau$ is $p_{\text{click}}(\tau)=\int
_{M_{0}-\lambda}^{M_{0}+\lambda}\left\vert b(k,\tau)\right\vert ^{2}dk$ .
Then, one performs a second measurement at the time $2\tau,$ and so on.
Finally, the no-click probability at the instant $t=n\tau$ is
$p_{\text{no-cklick}}(t=n\tau)=1-w_{\lambda}(\tau)\frac{1-p(\tau)^{n}%
}{1-p(\tau)}$ \cite{proccreta}. Being $w_{\lambda}(\tau\rightarrow0)=0,$ one
obtains a QZE (no-click). This result is still valid in the case
$E_{th}\rightarrow-\infty,$ $\Lambda\rightarrow\infty$ (when the free decay is
purely exponential), hence the QZE is purely induced by the detector. For the
link to continuos measurements, see \cite{shimizu,pascapulsed}. A side-effect
of Ref. \cite{gplast} is that the QZE obtained for bang-bang measurements and
the QZE through continuous measurements are in general different, hence one
could in principle check if the collapse of the wave function is a real
physical process as proposed in Ref. \cite{bassi}.

\textbf{Link to QFT \cite{zenoqft}: }In the framework of relativistic QFT one
obtains a picture similar to the QM\ case. In QFT, one uses the scalar fields
$S$ and $\varphi$ embedded in the Lagrangian \cite{zenoqft,duecan,lupo}:%
\begin{equation}
\mathcal{L}=\frac{1}{2}(\partial_{\mu}S)^{2}-\frac{1}{2}M_{0}^{2}S^{2}%
+\frac{1}{2}(\partial_{\mu}\varphi)^{2}-\frac{1}{2}m^{2}\varphi^{2}%
+gS\varphi^{2}\text{ .}%
\end{equation}
The decay process $S\rightarrow\varphi\varphi$ is analogous to the transition
$\left\vert S\right\rangle \rightarrow\left\vert k\right\rangle $ described
previously. Upon taking into account proper relativistic expressions (instead
of nonrelativistic ones), the QFT results can be obtained in the framework of
the Lee Hamiltonian by a due choice of the vertex function $f(k)$
\cite{duecan}. Hence, the non-exponential nature of decay and all the other
phenomena described in this section apply also for QFT (Eq. (\ref{aoft}) is
valid also in QFT). It would be interesting in the future to go beyond the use
of the Lee Hamiltonian's matching and evaluate the time evolution in QFT in
the interaction picture. While it is not expected to invalid previous results,
it would be anyhow an important theoretical achievement. This project is left
for the future.

\textbf{Decay of a moving particle \cite{time}: }Finally, we briefly comment
on the survival probability of a state with non-vanishing momentum $q$ (here,
we follow \cite{time}; for previous works, see Ref.
\cite{khalfin2,shirokov,stefanovich,urbano}). The survival probability
amplitude reads $a(t,q)=\int_{-\infty}^{\infty}dmd_{S}(m)e^{-i\sqrt
{m^{2}+q^{2}}t}$ (as in Eq. (\ref{aoft})) which implies that
$p(t,q)=\left\vert a(t,q)\right\vert ^{2}\neq p\left(  tM_{0}/\sqrt
{p^{2}+M_{0}^{2}}\right)  $, hence the usual Einstein's dilatation formula
does not hold \cite{shirokov,stefanovich,urbano}. In the exponential limit,
the non-decay (survival) probability reads $e^{-\Gamma_{q}t}$ with:
\begin{equation}
\Gamma_{q}=\sqrt{2}\sqrt{\left[  \left(  M_{0}^{2}-\frac{\Gamma^{2}}{4}%
+q^{2}\right)  ^{2}+M_{0}^{2}\Gamma^{2}\right]  ^{1/2}-\left(  M_{0}^{2}%
-\frac{\Gamma^{2}}{4}+q^{2}\right)  }\text{ }\label{gammap}%
\end{equation}
which \textit{differs} from the standard formula $\Gamma M_{0}/$ $\sqrt
{q^{2}+M_{0}^{2}}$. One can easily prove that for $\Gamma/M_{0}\ll1$, the
Einstein formula represents a very good approximation. Indeed, the maximal
deviation is obtained for $q_{\max}=\sqrt{2/3}M_{0}$, for which (normalized to
$M_{0}$) reads $\sim\left(  \Gamma/M\right)  ^{3}$. This is in almost all
cases a ridiculously small number. On the contrary, a boost transforms an
unstable state into its decay products. A boosted neutral pion is a two-photon state.

\section{Conclusions}

In this work we have reviewed some recent theoretical works on non-exponential
decay and the QZE which were not confirmed yet experimentally. The
non-exponential decay when two (or more) decay channels are present seems
promising. Others, such as the measurement of decay products soon after their
emission and the QZE induced by detectors are appealing but probably difficult
to measure. The measurement of deviations form the Einstein's formula is at
present not possible. Theoretically, the firm understanding of these issues in
QFT on a solid mathematical basis (and without using Lee Hamiltonian's
analogy) is an important outlook of the present work.

\textbf{Acknowledgments}: F. G. thanks financial support from the Polish
National Science Centre NCN through the OPUS project no. 2015/17/B/ST2/01625.


\end{document}